# Mixed-anion mixed-cation perovskite (FAPbI$_3$)$_{0.875}$(MAPbBr$_3$)$_{0.125}$: an *ab-initio* molecular dynamics study

Eduardo Menéndez-Proupin,*[a,b] Shivani Grover,[c] Ana L. Montero-Alejo,[d] Scott D. Midgley,[c] Keith T. Butler,[e] and Ricardo Grau-Crespo[c,*]

[a]*Departamento de Física Aplicada I, Escuela Politécnica Superior, Universidad de Sevilla, Seville E-41011, Spain.*

[b]*Departamento de Física, Facultad de Ciencias, Universidad de Chile, Casilla 653, Santiago, Chile.*

[c]*Department of Chemistry, University of Reading, Whiteknights, Reading RG6 6DX, UK.*

[d]*Departamento de Física, Facultad de Ciencias Naturales, Matemática y del Medio Ambiente (FCNMM), Universidad Tecnológica Metropolitana, José Pedro Alessandri 1242, Ñuñoa, Santiago 7800002, Chile.*

[e]*SciML, Scientific Computing Division, Rutherford Appleton Laboratory, Harwell, UK.*

* Corresponding authors. Emails: emenendez@us.es; r.grau-crespo@reading.ac.uk

Electronic Supplementary Information (ESI) available: atomic coordinates and additional figures.

Mixed-anion mixed-cation perovskites with (FAPbI$_3$)$_{1-x}$(MAPbBr$_3$)$_x$ composition have allowed record efficiencies in photovoltaic solar cells, but their atomic-scale behaviour is not well understood yet, in part because their theoretical modelling requires consideration of complex and interrelated dynamic and disordering effects. We present here an ab initio molecular dynamics investigation of the structural, thermodynamic, and electronic properties of the (FAPbI$_3$)$_{0.875}$(MAPbBr$_3$)$_{0.125}$ perovskite. A special quasi-random structure is proposed to mimic the disorder of both the molecular cations and the halide anions, in a stoichiometry that is close to that of one of today´s most efficient perovskite solar cells. We show that the rotation of the organic cations is more strongly hindered in the mixed structure in comparison with the pure compounds. Our analysis suggests that this mixed perovskite is thermodynamically stable against phase separation despite the endothermic mixing enthalpy, due to the large configurational entropy. The electronic properties are investigated by hybrid density functional calculations including spin-orbit coupling in carefully selected representative configurations extracted from the molecular dynamics. Our model, that is validated here against experimental information, provides a more sophisticated understanding of the interplay between dynamic and disordering effects in this important family of photovoltaic materials.

**Introduction**

Hybrid organic-inorganic halide perovskites (HOIHP) have allowed to develop a promising generation of solar cells. Initial breakthroughs were achieved using perovskite MAPbI$_3$ (MA=CH$_3$NH$_3$) as light absorber, reaching photo-conversion efficiencies (PCE) of 15% by 2013 [1]. In this material, the A, B, and X sites of the perovskite structure ABX$_3$ are occupied by the ions MA$^+$, Pb$_2^+$, and I$^-$, respectively. Other members of the HOIHP family are obtained when replacing iodide by other halide, replacing lead by another group IV cation, or replacing MA by another organic cation or caesium. The full family includes random alloys of the pure compounds. It soon turned out that MAPbI$_3$ is barely stable against decomposition into lead iodide and methylamine iodide, causing fast device degradation. There have been significant achievements in the stability by means of interface engineering [2-4], and by alloying the cations and halides [5]. In the so-called mixed-cation HOIHP, the perovskite A-site is randomly occupied by one of different cations, e.g, MA, HC(NH$_2$)$_2$ (formamidinium, FA) or the inorganic Cs. These mixed compounds show remarkable improvement in their stability and are related with the evolution of record PCE [3]. Some of the recent perovskite solar cell (PSC) record efficiencies have been obtained using light absorber perovskites with composition (FAPbI$_3$)$_{1-x}$(MAPbBr$_3$)$_x$ [6, 7], with $x \sim 0.15$, or with MAPbBr$_3$ as a trace [8]. This composition was introduced by Jeon et al. [9], showing that incorporation of MAPbBr$_3$ into FAPbI$_3$ stabilizes the perovskite phase of FAPbI$_3$ and improves the PCE, particularly for composition (FAPbI$_3$)$_{0.85}$(MAPbBr$_3$)$_{0.15}$.

The understanding of this material, and eventually the PSC development will benefit from atomic scale simulations of the bulk materials and the interfaces. The first obstacle is the lack of models for the randomly mixed compounds. Models of several binary alloys, with either a cation or lead substituted, have been published recently [10, 11]. In fact, some recent experimental studies of mixed-cation mixed-halide perovskites have been complemented with simulations with one element substituted [12]. In this work, an atomic scale model of the mixed-cation mixed-halide (FAPbI$_3$)$_{1-x}$(MAPbBr$_3$)$_x$ is presented. The mixing ratio $x = 0.125$ is considered for the model, which is close to the composition present in optimized PSC. The model is based on pseudo-random substitutions of FA and I in FAPbI$_3$, by MA and Br, respectively. We then study the structural and vibrational properties by means of ab initio molecular dynamics. The alloy properties are compared with those of the pure compounds FAPbI$_3$ and MAPbBr$_3$. The electronic structure of the alloy are also studied by means of first principles calculations on carefully selected representative configurations.

**Methods**

*Ab-initio* molecular dynamics (MD) simulations were performed to investigate the structural, dynamical, and electronic properties of (FAPbI$_3$)$_{0.875}$(MAPbBr$_3$)$_{0.125}$, as well as the end members of the solid solution FAPbI$_3$ and MAPbBr$_3$. Our MD simulations were conducted under NVT conditions at temperature $T$= 350 K using the CP2K package [13]. The ionic forces were calculated using first principles density functional theory (DFT). The hybrid Gaussian and plane wave method (GPW), as implemented in the QUICKSTEP module of the CP2K package [14], has been used. The forces for the molecular dynamics were calculated using the PBE functional [15] with the Grimme correction scheme DFT-D3 [16] to account for the dispersion interactions. The Kohn-Sham orbitals of valence electrons are expanded in a Gaussian basis set (DZVP-MOLOPT for Pb, I , Br, C, N, H) [17]. The effect of core electrons was included by means of dual space GTH pseudopotentials [18-20]. The functional minimizations were performed using the orbital transformation method [21, 22]. The timestep was set to 1 fs. A Nose-Hoover chain thermostat of length 3 was used, with a time constant of 10 fs during the initial 2000 steps, and 100 fs for the rest of equilibration and sampling stages. During the initial 2000 steps, each atom had one individual thermostat applied to; this is called a massive thermostat and facilitates a fast thermalization of systems with too different atomic masses, such as H and Pb. The simulations were extended up to 28, 26, and 23 ps for FAPbI$_3$, MAPbBr$_3$. and (FAPbI$_3$)$_{0.875}$(MAPbBr$_3$)$_{0.125}$, respectively. These systems were considered equilibrated after the initial 8, 6, and 5 ps, respectively, while sampling took place for the subsequent time ranges. The full set of atomic coordinates along the MD simulations are available in a public repository [23].

The electronic structure of selected configurations was calculated using the Vienna Ab Initio Simulation Package (VASP) [24], with the PBE functional with spin-orbit coupling (SOC), as well as with the hybrid functional PBE0 (with fraction of exact exchange modified to 0.188 and SOC as discussed in Ref [25]). SOC is mandatory to compute accurate conduction band energies in lead halide perovskites [26], while the hybrid functional is needed to obtain correct band gaps and band edge energies. Projected augmented wave (PAW) [27, 28] potentials of soft type were used for C, N and H, with 4, 5, and 1 valence electrons respectively. Standard PAW potentials have been used for Pb, Br, and I, which include 4, 7, and 7 electrons in the valence, respectively. A kinetic energy cutoff of 295 eV defines the plane waves basis set employed.

**Results and discussion**

**Model generation**

For FAPbI$_3$, we start from the structure of Ref. [10]. The 384-atom supercell was transformed into a 768-atom 4x4x4 cubic supercell by means of the lattice transformation $A' = A + B, B' = B - A$, where $A, B$ ($A', B'$) are the initial (final) supercell lattice vectors. The conventional cubic lattice parameter was set at the experimental value $a$=6.3620 Å [29]. For MAPbBr$_3$, we started from the structure of Ref. [30] with lattice parameter $a$=5.9328 Å. For our 768-atom supercell, $A' = B' = 4a$. The initial orientation of the MA cations is taken from the MAPbI$_3$ polymorphic model of Ref. [31].

The solid solution $(FAPbI_3)_{0.875}(MAPbBr_3)_{0.125}$ was modelled by means of a special quasi-random structure (SQS) [32], which was generated as follows. First, the set of all symmetrically different configurations within a 2x2x2 perovskite supercell with composition $FA_7MAPb_8Br_3I_{21}$, was obtained using the site occupation disorder method (SOD) [33]. For this purpose, cubic crystal symmetry (space group 221) of the parent lattice was assumed, considering FA and MA as point atoms. For each of the 62 inequivalent configurations, the number of Br atoms (out of eight halides) in the first coordination sphere around MA, is 0, 1, 2 or 3. Eight of these 2x2x2 supercells were then combined to form a 4x4x4 supercell configuration that satisfies two conditions: (i) Halide-halide pair correlation function is as close as possible to the expected one for the random distribution, i. e. $\langle S_i S_j \rangle = (2x-1)^2 = 0.5625$ [32], considering the fraction of halide sites occupied by Br, $x = 3/24$. For the best SQS configuration, the obtained pair correlation functions are 0.5625, 0.5070, and 0.3958 for the first, second and third coordination spheres, respectively. (ii) The distribution of MA-Br pairs is as close as possible to the binomial distribution expected for the random solution. The probability of finding $n$ Br ions as nearest neighbours to the MA (out of $N$=12 halides) is $P(n) = C(N,n)x^n(1-x)^{N-n}$. These numbers must by approximated to fractions $m/8$, as we combine eight 2x2x2 supercells to form the 4x4x4 supercell. This condition translates in the requirement of including two 2x2x2 configurations with $n$=0, three with $n$=1, two with $n$=2, and one with $n$=3, in the composition of the 4x4x4 supercell, as shown in Table S1 of ESI.

Finally, the point-like MA and FA cations were replaced in the obtained 4x4x4 supercell by the full set of atoms, giving them random initial orientation. The supercell size was set as $A = 4a$, with $a$=6.3115 Å, that results from linear interpolation of the experimental densities of the end compounds $FAPbI_3$ and $MAPbBr_3$. A theoretical lattice parameter could have been determined, but it needs expensive variable cell MD. Instead, some degree of validation was pursued by means of static variable cell optimization, enforcing the cubic shape of the supercell, and taking the initial atom coordinates from some random configuration of the MD. With this procedure, the optimized lattice parameter was found as 6.3125 Å, which is very close to the interpolated value used in the subsequent MD simulations with the NVT ensemble. Applying the same procedure to the end compounds, values 6.37253 Å, and 5.9312 Å, were obtained for $FAPbI_3$ and $MAPbBr_3$, respectively, which are also close to the experimental values above mentioned, used in our subsequent simulations.

**Structural analysis**

We now discuss the structural information derived from MD simulations. The descriptors of the local structure include pair distribution functions, angle distribution functions, and cation orientation, with especial attention given to variation induced by the cation and halide substitutions.

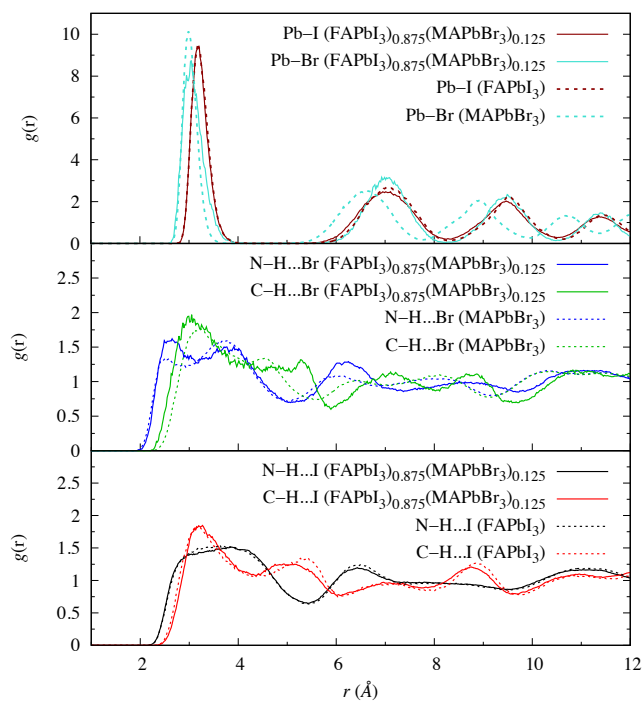

**Figure 1.** Partial pair distribution functions in $(FAPbI_3)_{0.875}(MAPbBr_3)_{0.125}$ and the pure compounds, for species Pb-I, Pb-Br and H-halide, differentiating the cases where H is bound to N or C.

**Figure 1** shows the partial pair distribution functions (PDFs) for the solid solution and the end members. Figure S1 in the ESI shows the same functions separately for the three compounds. The partial PDFs [34] are defined as:

$$g_{ij}(r) = \frac{V}{N_i N_j} \sum_k^{N_i} \sum_l^{N_j} \langle \delta(r - r_{kl}) \rangle, \quad (1)$$

where $N_i$ and $N_j$ are the number of atoms of the species $i$ and $j$ per volume $V$, containing $N$ atoms of all species. The brackets $\langle ... \rangle$ indicate the statistical average over the MD configurations. The total PDF is given as:

$$g(r) = \sum_{i,j} c_i c_j g_{ij}(r), \quad (2)$$

where the species fractional concentrations are $c_i = N_i/N$.

All the simulation supercells considered in this work contain 64 A-cations (MA or FA), 64 Pb, and 192 halide sites. In the simulation cell of $(FAPbI_3)_{0.875}(MAPbBr_3)_{0.125}$ there are 24 Br atoms, and eight MA. As seen in **Figure 1**, the Pb-Br PDF at the range of the first coordination shell in the solid solution has the maximum at a similar distance (3.06 Å bond length) to the corresponding in $MAPbBr_3$ (3.01 Å), whereas the Pb-I bond length is the same (3.20 Å) in $FAPbI_3$ and in the solid solution. The shorter Pb-Br bond distance compared with Pb-I implies that Pb-Br-Pb trios are stretched, and the Br moves less freely in the solid solution than in pure $MAPbBr_3$. **Figure 2** supports this statement, showing that the Pb-Br-Pb angles are closer to 180º in the mixture than in pure $MAPbBr_3$, while the Pb-I-Pb angles in the mixture tend to be lower than in $FAPbI_3$. Figure S2 in the ESI shows the same data in separate plots for both

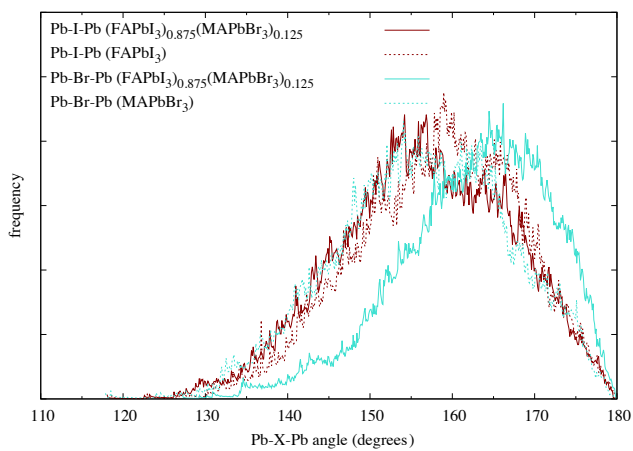

**Figure 2.** Distribution of Pb-X-Pb (X=Br, I) angles in the three compounds. Solid lines correspond to (FAPbI$_3$)$_{0.875}$(MAPbBr$_3$)$_{0.125}$, while dashed lines correspond to FAPbI$_3$ and MAPbBr$_3$.

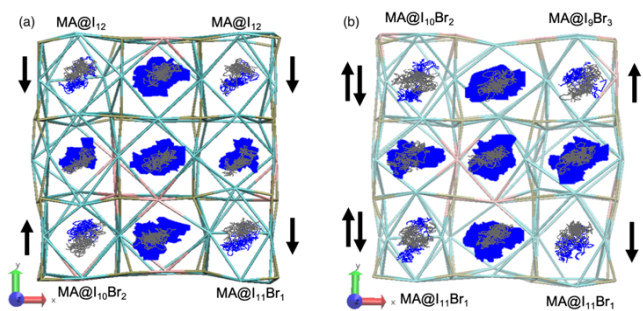

**Figure 3.** Density plots for N (blue) and C (grey) in a snapshot of the inorganic framework (Pb (ochre), I (cyan), Br (pink)) of the solid solution. Two different layers of the supercell are represented (a and b). Only the region that contains the MA cations is represented, corresponding to three lattice constants in each direction. The figure shows labels for the densities of the MA cations and the composition of the surrounding halogen atoms. The black arrows indicate the orientation of the MA dipoles within the cuboctahedra. The densities corresponding to the FA cations are not labelled.

Br and I angles, whereas Figure S3 shows the distribution of dihedral angles X-Pb-X-Pb for all compounds. The three distributions are centred at 0°, indicating that the dynamic structure is consistent with cubic perovskites. For a tetragonal phase HOIHP displaying the typical octahedra tilting, the dihedral angle distribution would be bimodal, with two maxima centred around a positive and a negative angle [35]. The heterogeneous distribution of the Pb-X-Pb angles of the solid solution (unlike that of pure compounds) must generate deformations in the halogen cages that surround the organic cations, together with disparate Pb-Br and Pb-I bond lengths within each octahedra. Since the generated SQS supercell has a relatively homogeneous distribution of Br atoms, it is expected that these deformations should be present throughout the solid solution. These deformations are expected to disturb in some way the dynamics of the rotations of the organic cations in the mixture, because the well-known coupling between rotations and the vibrations of the inorganic framework. Also, the rotation of organic cations is guided by their hydrogen bonds interactions with their environment.

In **Figure 1**, the H-X (X=Br, I) PDF marks the difference between the H bonded to N or C. First, there is the double peak structure of the (N-H--Br) PDF (Br with the H of the NH$_3$ group in MA) in pure MAPbBr$_3$. The maximum at 2.53 Å signals hydrogen bonds, while the maximum at 3.79 Å indicate non-bonding distances. For FAPbI$_3$, the PDF of N-H--I (with the H of NH$_2$ end groups in FA) presents a single broad peak. This difference between N-H--Br and N-H--I is kept in the solid solution, regardless of the cation which the N-H species belongs to. The C-H--X PDFs presents one single maximum at ~3Å. The maximum of the C-H--I PDF occurs at the same distance in FAPbI$_3$ and the solution. In contrast, the maximum of the C-H--Br PDF occurs at a slightly higher distance (compared with C-H--I PDF) in MAPbBr$_3$, but at a smaller distance in the solution.

The Pb-Br PDF first peak in the solid solution is broader than in pure MAPbBr$_3$. For higher coordination shells, the Pb-X PDF attain maxima at the same distances for both X=Br, I, although there are differences in the broadening and height of the PDFs. The Pb-halide distances for the second and higher coordination shells are determined by the lattice parameter. The decreased broadening of the Pb-Br PDF compared with Pb-I and with Pb-Br in MAPbBr$_3$ is difficult to interpret, but it may be related with the Pb-Br-Pb angles being closer to 180º, as shown in **Figure 2**.

The molecular dynamics shows that MA makes precession and rotations around its axis, with the NH$_3$ always pointing to the central Br (for 8 ps), although at the same time the CH$_3$ group points to other Br atoms. This can be seen in the video Br-1MA.mpg, provided in the dataset [23], which shows a local environment with 3 Br close to one MA. The Br atom at the centre is surrounded by 3 FA and 1 MA. The geometrical parameters are 4.0 Å in distance cutoff, and 30° in angle cutoff. The hydrogen bonds are not permanent but form and break dynamically at this temperature.

When looking at the density maps of the C and N atoms of the solid solution (**Figure 3**), it is evident that the MA cations have an almost unique orientation during the trajectory, which is apparent in the video. This fact seems independent of the number of Br atoms (0-3 atoms) close to the MA cation. Furthermore, the density maps show that the orientation of the NH$_3$ group in the MA is not always towards the Br atom of the cage, see the extreme bottom-right in **Figure 3b**. All the evidence shows that the MA cations in the solid solution have the rotation affected by two causes: the deformations of the halogen cage that surrounds them, and the favourable hydrogen bond interactions towards the Br atoms. These causes should also affect the rotation of the FA cations, but the N density maps of these cations appear almost spherical. For comparison, Ghosh et al [36] have shown a pronounced asymmetry in the N atom density map, together with stable cage deformation, in FA$_{0.9}$Cs$_{0.1}$PbI$_3$. In the next section some similarities with the case of FA$_{0.9}$Cs$_{0.1}$PbI$_3$ are exposed.

**Dynamic properties**

The orientation dynamics of organic cations is one of the fingerprints of HOIHP, modifying the dielectric response, carrier conductivity, and thermodynamic properties, among others. In MD, this information is available from vector autocorrelation functions [37] as follows

$$c(\tau) = \frac{1}{N}\sum_{n=1}^{N}\langle \hat{n}(t_n)\cdot \hat{n}(t_n+\tau)\rangle,$$

where $N$ is the number of different time origins averaged, the unit vector $\hat{n} = \boldsymbol{u}/|\boldsymbol{u}|$, and $\boldsymbol{u}$ is the relative position vector of pairs of atoms that define molecular orientation. **Figure 4** shows the functions $c(\tau)$ for three characteristic vectors: (i) vector C-N in MA$^+$ parallel to molecule dipole, (ii) vector C-H in FA$^+$, also parallel to molecule dipole and to the short molecule axis, and vector N-N in FA$^+$, associated to the so-called tumbling motion that rotates the long molecule axis [36]. It is seen that in the three cases, the curves decay more slowly in the solid solution (solid lines) than in the pure compounds (dashed lines), indicating that mixing slows the cation reorientation dynamics. In this sense, this is similar to the effect observed in FA$_{0.9}$Cs$_{0.1}$PbI$_3$ in Ref. [36].

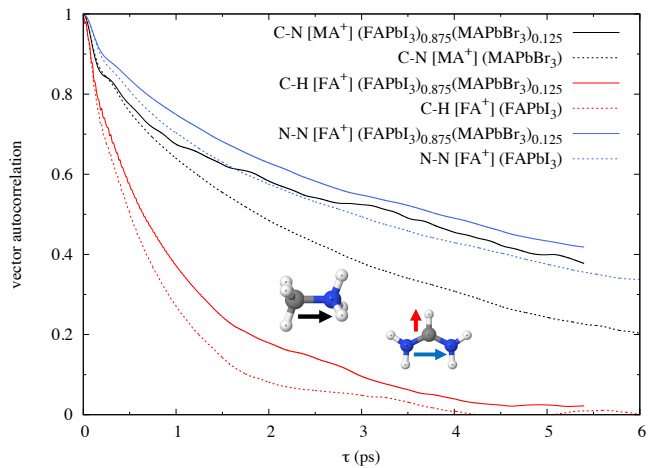

**Figure 4.** Vector autocorrelation of organic cations (rotation). The insets indicate the vectors in MA$^+$ (C-N), and FA$^+$ (C-H and N-N), using the same colours as in the corresponding curves.

Additional insight can be obtained from the evolution of the polar angle of vector C-N (MA$^+$) with respect to each Cartesian axis, displayed in figure S4. These cations are surrounded by 1, 2 or 3 Br atoms. It can be observed that proximity to 1 or 2 Br tend to fix the orientation of MA$^+$, and the events of orientation reversal occur rather abruptly. When MA$^+$ is surrounded by 0 or 3 Br, orientation reversals occur more frequently, and are not abrupt. Figure S5 in the ESI shows that in pure MAPbBr$_3$, MA$^+$ cations continuously change their orientation. The restriction of MA$^+$ motion in (FAPbI$_3$)$_{0.875}$(MAPbBr$_3$)$_{0.125}$ is contrary to the expectation that MA$^+$ could rotate more freely in a wider cuboctahedral cavity, given the larger lattice parameter, compared to MAPbBr$_3$.

These results on the reduced mobility of organic cations in the mixture are consistent with several experimental observations. Grüninger et al. [12] have performed an NMR study of the $^1H - ^1H$ correlations in (FA$_x$MA$_{1-x}$PbI$_3$) and (FAPbI$_3$)$_{0.85}$(MAPbBr$_3$)$_{0.15}$, their results suggesting a restriction of the mobility of the organic cations in (FAPbI$_3$)$_{0.85}$(MAPbBr$_3$)$_{0.15}$. Johnston et al [38] have arrived at a similar conclusion based on quasi-elastic neutron scattering measurements of triple-cation mixed halide perovskites in a low temperature hexagonal (non-perovskite) phase. They have found a correlation between cation rotation inhibition (particularly FA) and carrier lifetime, which is maximized when the Br content is 15%. They attributed the optimality of 15% Br content to that being the ratio at which approximately one every six halides are bromine. By means of Monte Carlo simulation they showed that for 15% Br content, the frequency distribution of Br is, from highest to smallest, 1, 0, 2, and 3 Br per unit cell.

Our observation that cation rotation is inhibited in proximity to 1 or 2 Br, even at this high temperature of 350 K, is qualitatively consistent with that frequency distribution. Simenas et al. (2021) [39] have applied dielectric spectroscopy to mixed-cation FA$_x$MA$_{1-x}$PbBr$_3$ and have obtained several activation barriers, concluding that FA presence raises the MA rotation barriers. They have also found several dielectric relaxation modes that are presence only in pure FAPbBr$_3$, suggesting that MA presence also hampers FA movements.

**Vibrational properties**

**Figure 5** shows the partial vibrational densities of states (VDOS) of the (FAPbI$_3$)$_{0.875}$(MAPbBr$_3$)$_{0.125}$, FAPbI$_3$, and MAPbBr$_3$. The vibrations of the inorganic backbone are in a band in the range 0-200 cm$^{-1}$. This band also contains important contributions from FA and MA. The band in the range 500-800 cm$^{-1}$ is due to FA vibrations only, while the band in the range 1000-1800 cm$^{-1}$ is due to both MA and FA vibrations. The band for frequencies larger than 3000 cm$^{-1}$ is due to the stretching of C-H and N-H bonds in the organic cations [40]. As can be appreciated from **Figure 5(a)** and **(b)**, the VDOS and the partial VDOS of the solid solution is almost identical to the weighted sum of the DOS of the pure compounds. This property may allow a method for quantification of the composition $x$ in the solution. In particular, the amount of FA$^+$ can be determined from isolated modes in the range 1675—1800 cm$^{-1}$, which are associated to the stretching of the double resonant C=N of FA$^+$. The out-of-plane N—H bending modes of FA$^+$, in the range 400—800 cm$^{-1}$ could also be used. The band of stretching modes should be more effective in practice, because there are no more modes at this high frequency. MA$^+$ modes display isolated bands or peaks in the ranges 900-980 cm$^{-1}$, 1235 cm$^{-1}$, and 1400-1500 cm$^{-1}$. The intensity ratios of some of these bands in IR spectra may allow a quantification of the MA/FA content. On the other hand, Br content cannot be easily determined because the Br and I partial VDOS overlap in the same spectral region.

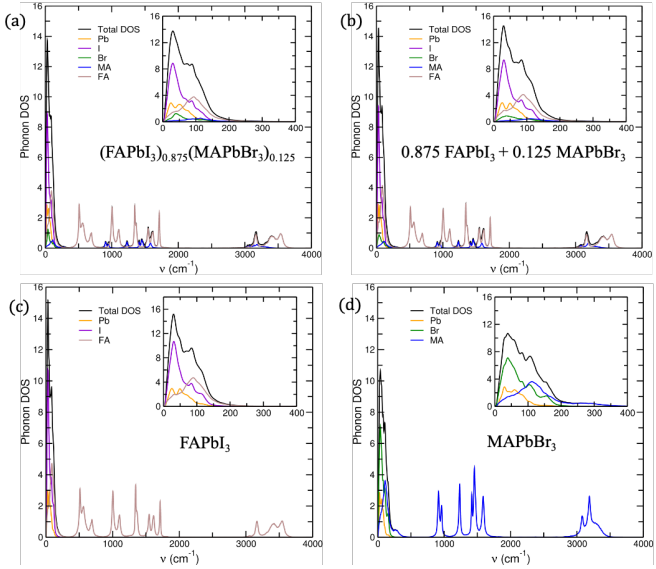

**Figure 5.** Phonon density of states and contribution of each crystal site to the total phonon density of states for (FAPbI$_3$)$_{0.875}$(MAPbBr$_3$)$_{0.125}$, FAPbI$_3$, MAPbBr$_3$, and the linear combination of VDOS 0.875 FAPbI$_3$ + 0.125 MAPbBr$_3$.

Moreover, the fact that VDOS of the solid solution is well approximated by the weighted VDOS of the compounds, suggests that contribution of vibrations to the free energy of mixing are relatively small and can be safely neglected in a first approximation to the free energy of mixing. This is important because it allows to analyse the stability of the solid solution taking into account only energies and configurational entropic effects, as we do in the next section.

**Thermodynamics of mixing**

We have calculated the enthalpy of mixing per formula unit of (FAPbI$_3$)$_{0.875}$(MAPbBr$_3$)$_{0.125}$ with respect to the pure compounds FAPbI$_3$ and MAPbBr$_3$ using the equation:

$$\Delta H_{\mathrm{mix}} = \frac{1}{64}\{E[(\mathrm{FAPbI}_3)_{1-x}(\mathrm{MAPbBr}_3)_x] - (1-x)E[\mathrm{FAPbI}_3] - xE[MAPbBr_3]\}$$

where the $E$ values are the average energies for the corresponding supercells from the ab initio MD, and $x$=0.125 is the MAPbBr$_3$ molar fraction. We obtain $\Delta H_{\mathrm{mix}}$= 1.39 kJ/mol, which means that the mixing is endothermic. However, the configurational entropy

$$S_{\mathrm{conf}} = -4k_B(x\ln x + (1-x)\ln(1-x))$$

is large (12.5 J/mol K for $x$=0.125) due to the factor of 4, which appears because there are four mixed sites (1 MA/FA and 3 Br/I) per formula unit. Assuming that the enthalpy of mixing and the configurational entropy contribution are the dominant effects on the stability, the free energy of mixing is predicted to be negative ($\Delta G_{\mathrm{mix}}$ = -3.00 kJ/mol).

To assess whether the mixed perovskite can be expected to be stable with respect to phase separation into MAPbBr$_3$-rich and FAPbI$_3$-rich compounds, it is not enough to obtain the mixing free energy at the given composition (which quantifies the stability with respect to the pure phases). We also need to calculate the free energy at the whole range of compositions to assess if the given composition lies within a miscibility gap. Ab initio molecular dynamics are too computationally expensive to perform for the whole range of compositions, but we can extrapolate our one-point result using the widely used regular solid solution model [41-44], where the enthalpy of mixing is given by:

$$\Delta H_{\mathrm{mix}} = W_0 x(1-x)$$

for which we obtain $W_0$=12.7 kJ/mol. With this enthalpy parameter, the mixing free energy in the regular solution model is convex for the full range of compositions at 350 K, implying that there is no miscibility gap: with 4 mixed sites per formula unit, the enthalpy parameter $W_0$ would need to be greater than the critical value of 8$k_B T$, or ~25 kJ/mol at 350 K, for a miscibility gap to exist.

The above analysis suggests that the mixed perovskite is indeed thermodynamically stable against phase separation. Although we have ignored other possible contributions to the thermodynamics of mixing (including vibrational and rotational contributions to the entropy of mixing), the low enthalpy of mixing in the regular model, well below the critical value for miscibility gaps, suggests that our prediction is not likely to change by including such contributions.

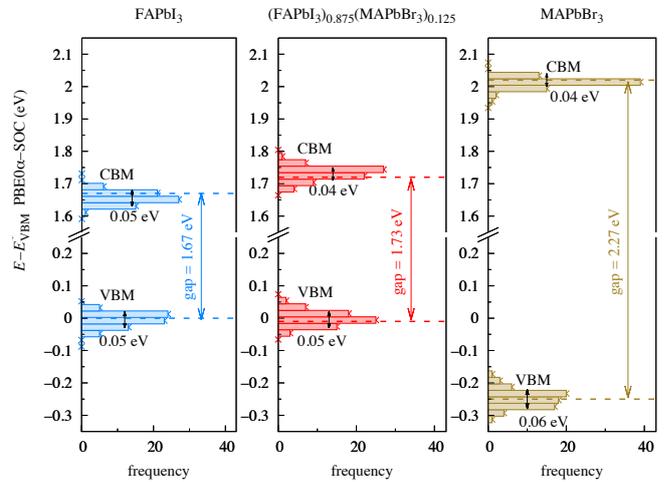

**Figure 6.** Distribution of energy levels of the mixture and the pure compounds at hybrid PBE0($\alpha = 0.188$)-SOC. $E^*_{VBM}$ is the VBM state energy of the FAPbI$_3$ selected configuration. The dashed lines represent the energies of the VBM and CBM states for each selected configuration, and the corresponding gap is included. The black double arrows indicate FWHM, the values are added for each distribution.

**Electronic structure**

To compare the energy spectra of the two pure compounds and the solid solution, the energy scales must be aligned with a well-defined procedure. The energy zero in VASP calculations is the average electrostatic potential, which includes the local part of the PAW potentials. As the three compounds have different compositions, the energy scales have different references. In order to put the electronic structures in the same reference, we have applied constant shifts to the electron energies of (FAPbI$_3$)$_{0.875}$(MAPbBr$_3$)$_{0.125}$ and MAPbBr$_3$ so that the Pb core levels are aligned.

Table 1 shows the average electrostatic potential (times the electron charge) computed in small spheres around the Pb nuclei. The fluctuations of this potential mimic the variations of core level shifts with the atomic environment [45]. Table 1 shows the mean values and the standard deviation of the sphere-averaged potentials. For the mixture (FAPbI$_3$)$_{0.875}$(MAPbBr$_3$)$_{0.125}$ these parameters were calculated separately as a function of the number of Br atoms in the corners of the PbI$_{6-n}$Br$_n$ octahedra. The variation between octahedra with 0 and 3 Br, is 0.20 eV, while the standard deviation of the full set of Pb is similar to the one in the pure compounds. This suggests that core-level measurements (say, using X-ray photoemission spectroscopy) would not show increased broadening of Pb core levels compared to the pure compounds, and therefore that the Pb core levels can be regarded as a common energy reference for the alloy and the pure compounds. For the Pb core levels to be aligned in the three compounds, the energy scales must be shifted by 0.12 and 1.00 eV for (FAPbI$_3$)$_{0.875}$(MAPbBr$_3$)$_{0.125}$ and MAPbBr$_3$, respectively. The same shifts modify all single-electron energies and have been applied in the DOS plots below.

**Table 1.** Average electrostatic potential in spheres around Pb atoms in the studied compounds.

| Average electrostatic potential | Pb | Mean (eV) | Std. dev. (eV) | Correction shifts_to_align (eV) |
|---|---|---|---|---|
| (FAPbI$_3$)$_{0.875}$(MAPbBr$_3$)$_{0.125}$ | PbI$_6$ | -45.97 | 0.15 | |
| | PbI$_5$Br | -46.05 | 0.14 | |
| | PbI$_4$Br$_2$ | -46.09 | 0.16 | |
| | PbI$_3$Br$_3$ | -46.17 | 0.20 | |
| | Pb$_{all}$ | -46.04 | 0.16 | 0.12 |
| FAPbI$_3$ | Pb$_{all}$ | -45.92 | 0.16 | 0 |
| MAPbBr$_3$ | Pb$_{all}$ | -46.92 | 0.18 | 1.00 |

The valence band maximum (VBM) of each compound has been determined as the average energy of the higher occupied crystal orbital (HOCO), taken from a subset of 70 configurations of the MD ensembles, separated by 250 fs. **Figure 6** shows the distribution of the HOCO energies for the three compounds. The energy dispersion has been computed at PBE-SOC level. The same procedure has been applied to obtain the conduction band minimum (CBM) as the average of the lowest unoccupied crystal orbitals (LUCO). Moreover, for each compound we have identified one configuration that have the HOCO and LUCO energies very close to the average VBM and CBM. For these special (selected) configurations (see coordinates in ESI) we have calculated the electronic structure using the hybrid functional PBE0(0.188)-SOC, as well as with PBE-SOC. From these calculations with the representative configurations, the differences between PBE0(0.188)-SOC and PBE-SOC have been computed for the VBM and the CBM and have been applied as corrections to the set of HOCO and LUCO energies in **Figure 6**. Also Figure S6 in the ESI shows the DOS for the mixed perovskite and the pure compounds, computed with the PBE functional. Each plot includes the DOS of the special configuration. All energies are referred to the VBM of FAPbI$_3$, which is defined as the average HOCO, as explained above. The energy scales of (FAPbI$_3$)$_{0.875}$(MAPbBr$_3$)$_{0.125}$ and MAPbBr$_3$ are shifted to align the average Pb core level energies with that in FAPbI$_3$. As the PBE functional is well known to underestimate the bandgap, the electronic structure of the selected representative configuration of each compound was recalculated with the hybrid functional PBE0(0.188)-SOC. The DOS and PDOS calculated with this functional are shown in **Figure 7** and S7.

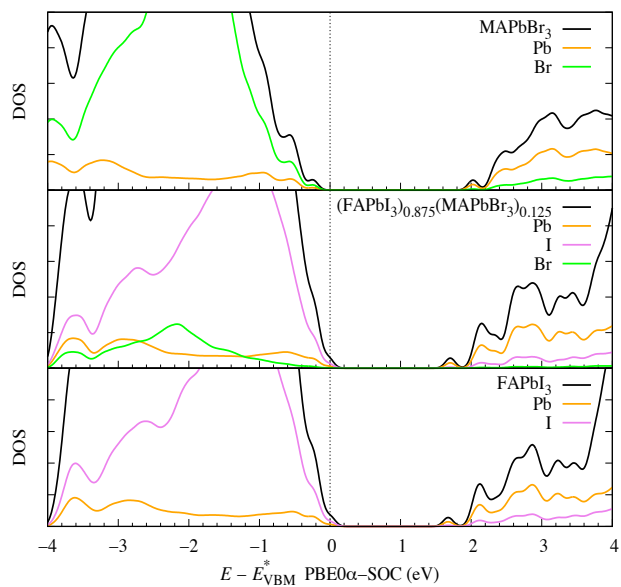

**Figure 7.** DOS and PDOS of the solid solution and the pure compounds computed at hybrid PBE0($\alpha = 0.188$)-SOC. $E^*_{VBM}$ is the VBM state energy of the FAPbI$_3$ selected configuration. The energy scales of (FAPbI$_3$)$_{0.875}$(MAPbBr$_3$)$_{0.125}$ and MAPbBr$_3$ are shifted to align the average Pb core level energies with that in FAPbI$_3$.

With the same approach, the imaginary part of the dielectric function was computed, which is shown in Fig. S8. The dielectric function was computed within the random phase approximation (RPA), with the longitudinal gauge, and using only the gamma point wavefunctions. One shortcoming of the RPA is that it cannot account for exciton effects. Moreover, the rough k-point sampling that we can afford in the calculation of the dielectric function does not allow to compare the theoretical and experimental dielectric functions. However, the theoretical calculation allows to confirm that the VBM to CBM transitions are dipole allowed. Therefore, one can conclude that the optical gaps almost coincide with the bandgaps so long as exciton binding energies are negligible.

**Table 2** shows the theoretical bandgaps, computed with both hybrid functionals, as well as the experimental optical gaps. The experimental optical gaps of FAPbI$_3$ and (FAPbI$_3$)$_{0.875}$(MAPbBr$_3$)$_{0.125}$ were obtained by means of the

Tauc's plots, shown in Figure S9, as well as from the literature. It is apparent that the PBE0(0.188)-SOC method provides a band gap in better agreement with the experimental optical gaps than HSE06-SOC. Let us recall that the mixing fraction 0.188 was obtained for MAPbI$_3$ [25], therefore it is not expected to work so well for different compositions, but it is still more accurate than the standard HSE06-SOC for these HOIHP perovskites. Moreover, both functionals reproduce well the trend that the gap of the mixture is 0.06 eV larger than the gap of FAPbI$_3$.

**Table 2.** Comparison of experimental and theoretical gaps.

| Compound | Gap energy (eV) | | |
|---|---|---|---|
| | Experimental (optical) | Theoretical (PBE0(0.188)-SOC) | Theoretical (HSE06-SOC) |
| FAPbI$_3$ | 1.53[a] 1.52[b] | 1.67 | 1.35 |
| (FAPbI$_3$)$_{0.875}$(MAPbBr$_3$)$_{0.125}$ | 1.59[b] | 1.73 | 1.41 |
| (FAPbI$_3$)$_{0.85}$(MAPbBr$_3$)$_{0.15}$ | 1.58[a] 1.61[b] | | |
| MAPbBr$_3$ | 2.31[b] | 2.27 | 1.96 |

[a] From Tauc's plots, Figure S9.
[b] From Ono et al [5, 46].

The theoretical bandgaps of (FAPbI$_3$)$_{0.875}$(MAPbBr$_3$)$_{0.125}$ shown in Table 2 are close to the values predicted by the Vegard law, 1.75 or 1.43 eV with PBE0(0.188)-SOC or HSE06-SOC, respectively. This is consistent with experimental data summarized by the equation:

$$E_g(x,y) = 1.58 + 0.436x - 0.058y + 0.294x^2 + 0.0199xy \quad (3)$$

where the non-linear term for $x = 1 - y = 0.125$ contributes only 4 meV. Eq. (3) was obtained by Jacobsson et al. [46], by fitting the data of 49 synthetized mixtures that span the compositional space of FA$_y$MA$_{1-y}$Pb(I$_{1-x}$Br$_x$)$_3$.

There is no significant difference in the imaginary part of the dielectric functions of FAPbI$_3$ and (FAPbI$_3$)$_{0.875}$(MAPbBr$_3$)$_{0.125}$, as shown in Figure S8 of the ESI. This is interesting because Brauer et al. [47] have recently reported a large difference in the measured attenuation coefficients of MAPbI$_3$ and (FAPbI$_3$)$_{0.8}$(MAPbBr$_3$)$_{0.2}$, the latter being much higher in the range 2.5-3.1 eV. They have correlated this feature with the increased carrier lifetime in the mixed perovskite. From our calculation it seems that FAPbI$_3$ and (FAPbI$_3$)$_{0.8}$(MAPbBr$_3$)$_{0.2}$ are similar in this respect.

**Conclusions**

An atomic scale model of the random solution (FAPbI$_3$)$_{0.875}$(MAPbBr$_3$)$_{0.125}$ has been proposed. Several computed properties have been shown to reproduce available experimental measurements, such as bandgaps, lattice constant, cation orientation dynamics, and thermodynamic stability. Other physical properties are reported, including distance and angle distribution functions, vibrational and other dynamical properties, as well as the electronic structure. All these properties of the solid solution (FAPbI$_3$)$_{0.875}$(MAPbBr$_3$)$_{0.125}$ are compared with those of the end compounds MAPbBr$_3$ and FAPbI$_3$. Our MD simulations at 350 K, show for (FAPbI$_3$)$_{0.875}$(MAPbBr$_3$)$_{0.125}$ a certain locking effect on the orientation of the organic cations, compared with the pure compounds. Thermodynamic calculations suggest there is no miscibility gap for the random alloying of MAPbBr$_3$ and FAPbI$_3$. Our model is therefore useful for the detailed understanding of the physical behaviour of this important material, accounting for both disorder and dynamic effects, and will allow future investigation of other bulk properties as well as of the behaviour of its surfaces and interfaces.

**Author Contributions**

This work was conceived by EMP and RGC. The SQS model was designed by RGC and EMP. MD simulations were made by SDM, SG and EMP. Analysis of the MD was done by SG, ALMA, and EMP. The electronic structure calculations were done by ALMA. The thermodynamic analysis was done by SG and RGC. Most of the pictures were composed by SG and ALMA. KTB contributed to the interpretation and discussion of the results. The first draft was written by EMP and RGC. All authors contributed to the final manuscript.

**Conflicts of interest**

There are no conflicts to declare.

**Acknowledgements**

Powered@NLHPC: This research was partially supported by the supercomputing infrastructure of the NLHPC (ECM-02). It made use of ARCHER, the UK's national high-performance computing service, via the UK's HPC Materials Chemistry Consortium, which is funded by EPSRC (EP/R029431). We are also grateful to the UK Materials and Molecular Modelling Hub for computational resources, which is partially funded by EPSRC (EP/P020194/1 and EP/T022213/1). EMP acknowledges support by the ANID/CONICYT/FONDECYT Regular 1171807 grant, and the University of Reading for its hospitality. SG thanks the Felix Trust for a postgraduate studentship. ALMA acknowledges ANID/CONICYT/FONDECYT Iniciación 11180984 grant. We also thank Xin-Gang Zhao for sharing the polymorphic structure of MAPbI$_3$ of Ref. [31].